\documentclass[preprint, prl]{revtex4}%
\usepackage{graphicx}%
\usepackage{amsmath}%
\usepackage{amsfonts}%
\usepackage{amssymb}
\renewcommand\Re{\operatorname{Re}}
\renewcommand\Im{\operatorname{Im}}

\begin{document}

\title{On Quantum Tunneling in Real Time}

\author{Neil Turok}

\affiliation{Perimeter Institute for Theoretical Physics, Waterloo, ON N2L 2Y5, Canada}

\begin{abstract}
A detailed real time description of quantum tunneling in the semiclassical limit is given, using complex classical trajectories. This picture connects naturally with the ideas of post-selection and weak measurement introduced by Aharonov and collaborators. I show that one can precisely identify the {\it complex} classical trajectory which a post-selected  tunneling particle has followed, and which dominates the path integral in the limit as Planck's constant $\hbar$ tends to zero. Detailed analytical calculations are presented for tunneling in cubic and quartic potentials. For a long post-selected tunneling time, the imaginary part of the tunneling coordinate is found to achieve very large values just before the particle tunnels. I discuss how the real and imaginary parts of the particle's coordinate may, in principle, be independently measured using weak measurements.  It would be  very interesting to observe this effect, which would  demonstrate the essential role of complex numbers in our closest possible classical description of reality.  Extensions to quantum field theory and general relativity are briefly discussed. 
\end{abstract}

\maketitle

Quantum tunneling is one of the most important and universal phenomena distinguishing the quantum and classical pictures of the world. Whereas a classical system can become `stuck'  in a confined region of phase space, the corresponding quantum system is typically able to explore the whole of it. In tunneling phenomena, quantum mechanics makes possible what is classically impossible. 

We are used to thinking of reality as described by real numbers - the spacetime coordinates, momenta and energies of particles for example. But in quantum physics, the imaginary number $i$ enters physics in a central way. Observable quantities such as those just mentioned are associated with Hermitian operators, and the formalism of quantum mechanics guarantees that measurement of such quantities, in the usual sense, can yield only real numbers. However,  such measurements - called strong measurements - interfere drastically with the state of the system being measured, in a sense `shoe-horning' it into a set of real values defined by the measuring apparatus. It is natural to ask whether a more gentle probe of reality such as weak measurements \cite{adv,ah80} might somehow directly reveal the role of complex numbers. In this paper, I study quantum tunneling as an interesting arena where such a test may be performed. A preliminary account of this work was given in \cite{ntorig}. 

Quantum tunneling is often exponentially suppressed, with the exponent being inversely proportional to Planck's constant $\hbar$. This is most easily seen via  Feynman's path integral: the wavefunction for a particle at time $t_f$ is given by the path integral
\begin{equation}
\Psi(x_f,t_f)={\cal N} \int Dx \int dx_i e^{{i\over \hbar}S (x_f,t_f,x_i,t_i) }\Psi(x_i,t_i).
\label{e1}
\end{equation}
In the limit as $\hbar$ is taken to zero, one expects the integral to be dominated by stationary points of the exponent, {\it i.e.}, solutions $x_{Cl}(t) $ of the classical equations of motion. An important difference with classical mechanics, however, is that the solutions must satisfy boundary conditions set at the initial and final times. For example, in (\ref{e1}), the initial condition for the classical solution is obtained by varying the   integrand (including the initial wavefunction) with respect to $x_i$, whereas the final condition is just $x_{Cl}(t_f)=x_f$. More generally, the final condition is similarly determined by including the final wavefunction, integrated over $x_f$, and varying the integrand with respect to $x_f$.
Notice that the dominant classical path or paths are only determined once both the initial and final states are specified. Aharanov and collaborators~\cite{adv, ah80} have already pointed out this natural occurrence of `pre- and post-selection' in the formalism of quantum mechanics and emphasized how it distinguishes quantum from classical mechanics.  Here I am pointing out that both selections are necessary in order to determine the dominant classical path or paths in the semi-classical limit.  Notice that there is absolutely no requirement that the dominant path or paths be real: in general they will be complex. Nevertheless, as $\hbar$ is sent to zero, this complex path (or set of paths) will accurately describe the behavior of the system and its interaction with other systems. 

In previous work, Carl Bender and collaborators noticed that complex classical solutions exhibit many of the properties normally associated with quantum tunneling, and numerically explored solutions very similar to those studied here. In particular, they noticed a key point: that solutions with real energy never tunnel and that the imaginary part of the energy is {\it essential} to describing tunneling. However, because they did not appreciate the connection with Feynman's path integral they were unable to discuss or derive the relevant real-time boundary conditions. Hence, in their work, the relationship between quantum mechanics and complex classical solutions remained somewhat mysterious, leading them to state, for example in Ref.~\cite{bendera}, that ``the ideas discussed here might be viewed as a vague alternative version of a hidden-variable formulation of quantum mechanics," and in Ref.\cite{benderb}, to claim that quantum tunneling is a kind of ``anomaly" exhibited by classical mechanics in the limit that the imaginary part of the energy tends to zero. The present work, to the contrary, {\it derives} the complex classical solutions as saddle points to Feynman's path integral in the semi-classical ($\hbar\rightarrow 0$) limit and, hence, removes any mystery as to their origin or meaning. Furthermore, this more precise (and conservative) interpretation, and the connection with post-selection, allows me to propose a weak measurement through which the complex nature of the tunneling solutions may be experimentally confirmed. In future work, I shall extend the quantum-complex classical correspondence to interference phenomena, which occur when more than one classical solution contributes in the $\hbar\rightarrow 0$ limit.

As a first example, consider a particle with action 
\begin{equation}
S = \int dt \left({1\over 2} m \dot{x}^2-V(x)\right), \, V(x)={1\over 2} \kappa x^2-{1\over 2} \lambda x^4.
\label{e2}
\end{equation}
It is convenient to rescale $t \rightarrow \sqrt{m/\kappa}\, t$ and $x\rightarrow  \sqrt{\kappa/\lambda} \,x$, {\it i.e.}, to measure time in units of the `false vacuum' oscillation period and $x$ in units of the location where $V$ crosses zero. Thus we obtain $S =( \kappa^{3\over 2} m^{1\over 2}/ \lambda) \int dt {1\over 2} (\dot{x}^2-x^2+x^4)$. The standard approach to semiclassical tunneling \cite{Coleman:1977py} is based on continuing the classical zero-energy solution, $x(t) = -1/\sin(t)$, to imaginary time $\tau= i (t+{\pi\over 2})$. This yields the Euclidean `bounce' solution $x(\tau)=  1/\cosh(\tau)$. For $-\infty<\tau<0$, this solution interpolates between the false vacuum, $x=0$, and the point $x=1$ at which the particle emerges from under the barrier (at zero energy) and commences its downhill roll to infinity. The exponent in the tunneling rate is given by squaring the wavefunction: one finds $ i (S-S^*)/\hbar = -S_{E}/\hbar = -{2\over 3} ( \kappa^{3\over 2} m^{1\over 2}/ \hbar \lambda),$ where the Euclidean action $S_E$ for the `bounce' is evaluated over $-\infty <\tau <\infty$. Whilst powerful, the Euclidean approach to tunneling has certain deficiencies. The dependence on the initial state is very implicit, and it is difficult to answer real-time questions, such as `where is the particle most likely to be found at each moment of time, as it tunnels?' and, more directly, `how {\it did} the particle get through the barrier?'

As we have explained, with pre- and post-selection, we can generally expect that tunneling will be described by one or more complex classical solutions. Assume, for example, that the system was initially prepared in a Gaussian localized around $x_i=0$, $\Psi(x_i,t_i)\sim e^{-x_i^2/4 L^2}$. Inserting this into (\ref{e1}) and stationarizing the integrand's exponent, from the Hamilton-Jacobi equation $p_i=-(\partial S /\partial x_i),$ we find the boundary condition $x_i/L+ p_i/\Delta p=0$, with $\Delta p= \hbar/(2 L)$. For the `false vacuum ground state' we have $L=1/\sqrt{2}$ in our dimensionless units, and we obtain the following mixed boundary conditions for the classical solution:
\begin{equation}
x+i\dot{x}=0\,\, {\rm at}\, \,t=t_i;  \qquad x=x_f \,\,{\rm at} \,\, t=t_f.
\label{e3}
\end{equation}
The first  condition says that the solution for $x$ is initially `pure positive frequency';  the second condition just says that it must attain the final real position $x_f$ at the final time.

The general solution to the classical equations of motion following from the action (\ref{e2}) involves  two arbitrary integration constants: the energy $E$ and a time delay $t_0$, both of which may be complex. Equations (\ref{e3}) provide two complex equations which fix these two unknowns. As we shall see, in the solutions which describe quantum tunneling, both $E$ and $t_0$ possess small imaginary parts. 

\begin{center}
\includegraphics[height=4.5in] {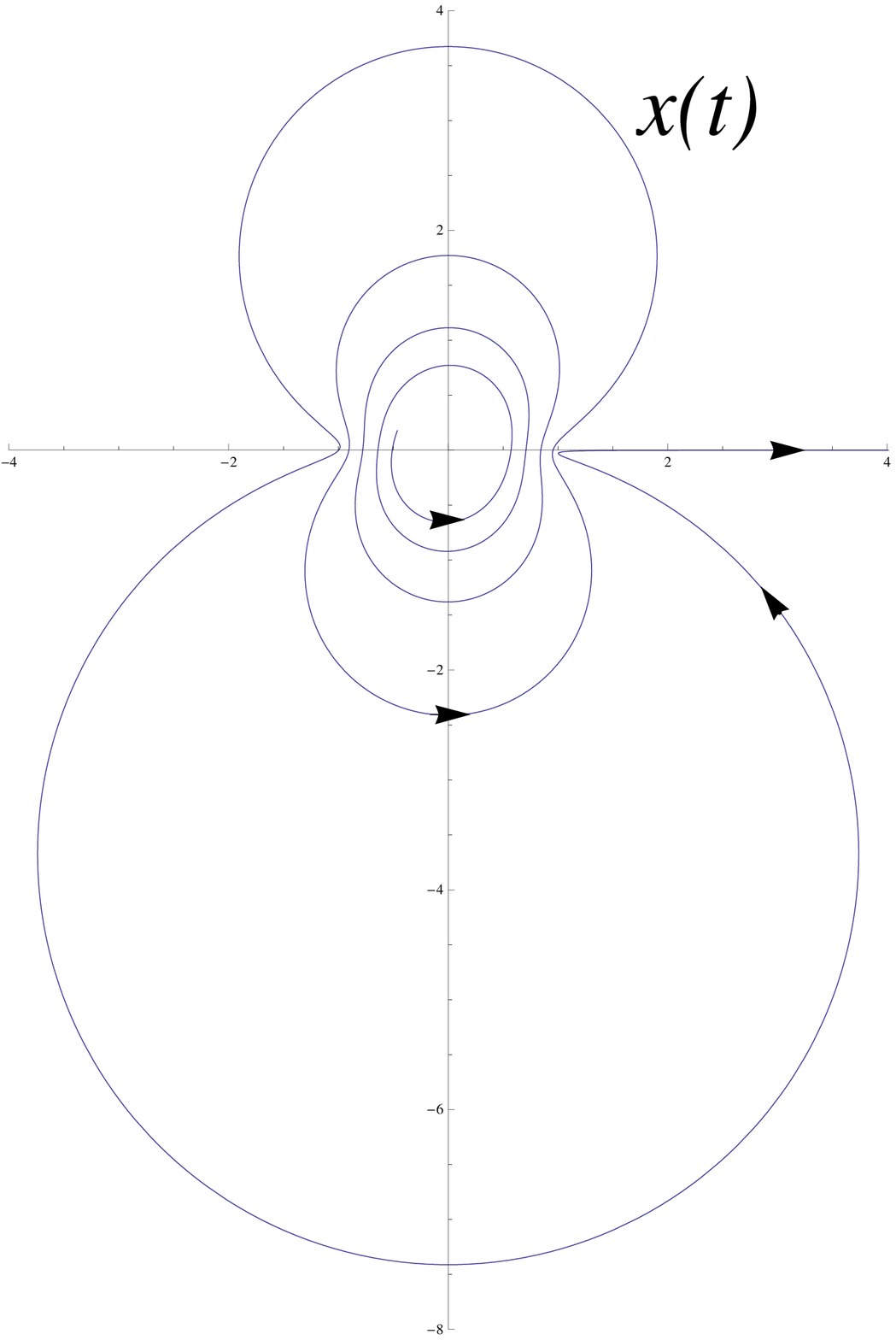}
\label{t1.eps}%
\end{center}
\begin{footnotesize}
Figure 1. For the potential  $V(x)= {1\over 2} (x^2- x^4)$, a complex classical solution $x(t)$ satisfying the boundary conditions in (\ref{e3}) with $t_i=-30$, $t_f=0$, $x_f=\infty$  is plotted for real times $t_i <t <t_f$  in the complex $x$-plane. Notice that the particle tunnels by going {\it around} the barrier in the complex $x$-plane, approaching the barrier from the  {\it outside} before bouncing off it and rolling to infinity. The energy of this solution is $-.003+.03 i$.
\end{footnotesize}

For the action (\ref{e2}), the general classical solution is
\begin{equation}
x_{Cl}(t)={1\over \sqrt{1+m}} {1\over {\rm sn} ( {t_0-t\over \sqrt{1+m}} |m)}, \quad E={m\over 2 (1+m)^2}
\label{e4}
\end{equation}
where ${\rm sn}(u|m) $ signifies the Jacobi elliptic function with argument $u$ and parameter $m$. As $m\rightarrow 0$ we find the `bounce' solution given above. 
The Jacobi elliptic functions are meromorphic and doubly periodic in the complex $u$-plane, with the period determined by $K(m) \equiv  \int_0^{\pi/2} dt  (1-m (\sin t)^2)^{-1/2}$, a quantity known as the `quarter-period.' Then  for any complex $m$, $2 K(m)$ and $i K'=i K(1-m)$ define two sides of a parallelogram which constitutes the fundamental domain of the function in the complex $u$-plane. For the sn function, two adjacent corners of this parallelogram, separated by $2 K(m)$,  are zeros and the other two, also separated by $2 K(m)$, are simple poles. Adjacent poles and zeros are separated by $iK'$.

The classical saddle point solution must satisfy the two boundary conditions in (\ref{e3}), which fix the two complex numbers $E$ and $t_0$.  The final condition becomes simple when $x_f\gg1$. Near the pole, we may express $x(t_f)$ as a Laurent series $x\approx (t_0-t_f)^{-1}$ plus regular terms. Inverting this series, we find $t_0-t_f=x_f^{-1}+o( x_f^{-3})$. Thus,  $t_0\rightarrow t_f$ as $x_f\rightarrow \infty$ so that, for large $x_f$, the first integration constant is uniquely fixed.

The initial condition in (\ref{e3}) may then be used to fix the energy $E$ or, equivalently, the parameter $m$.  Let us assume $T\equiv t_f-t_i \gg 1$, so the particle takes many false-vacuum oscillation periods to tunnel. For small $m$,  we have $K\approx {\pi\over 2}(1+{m\over 4})$, $K'\approx-{1\over 2} \ln (m/ 16)$ and the `nome' $q=e^{-\pi K'/K} \approx m/16$ is small, so we can express the solution using the Lambert series, 
\begin{equation}
{1\over {\rm sn}(u|m)} ={\pi\over 2 K}  \left( {1\over \sin U } +\sum_0^\infty {4 q^{2n+1}\over 1-q^{2n+1}} \sin (2n+1)U\right)
\label{e5}
\end{equation}
where $U={\pi \over 2 K} u$. Using this to re-express (\ref{e4}), we find $U=A(t_0-t)$, $A=\pi /(2 K  \sqrt{1+m}) \approx1-3m/4$. If $m$ has a small, positive imaginary part, then $Z\equiv e^{i U} $ becomes large at early times. We can then further expand (\ref{e5}) in inverse powers of $Z$ and in  $m$ to obtain
\begin{equation}
x(t)\approx 2 i Z^{-1} + 2 i Z^{-3}+{m\over 8 i} Z+\dots.
\label{e6}
\end{equation}
From the initial condition (\ref{e3}), we have $0=x+i\dot{x}\approx -4 i Z^{-3}+{m\over 4 i} Z+\dots$. Hence $Z$ is of order $m^{-{1\over 4}}$ at $ t_i$. The neglected terms in the initial condition and in (\ref{e6}) are then seen to be suppressed by $O(m^{1\over 2})$ relative to those kept, which justifies the expansion. Setting $m=i \epsilon$, we obtain a transcendental equation for $\epsilon$,
\begin{equation}
3 \epsilon T e^{3 \epsilon T}\approx 48iT e^{-4 i T},
\label{e7}
\end{equation}
which may be solved in terms of the Lambert W function, also called the product logarithm. The solutions are given by $\epsilon= {1\over 3} T^{-1} W_n(48iT e^{-4 i T})$, where $n$ is an integer which labels the branch of the Lambert W function: $n=0$ is the principal branch, for which, very roughly, $\Re(\epsilon) \sim {1\over 3} T^{-1} \ln T $ and $\Im(\epsilon)\sim O(1/T)$.  The $W_n$ function is tabulated, for example, in Mathematica: for the higher branches, the imaginary part of $\epsilon T$ has a progressively greater magnitude. In order to determine which of these solutions dominate in the semiclassical limit, we calculate the exponent appearing in the path integral, including the contribution from the initial wavefunction, in the same approximations used above. The result for the probability exponent is \cite{Footcorr}:
\begin{equation}
i{(S-S^*)\over \hbar} \approx {\kappa^{3\over 2}m^{1\over 2} \over \hbar \lambda}  \left( -{2\over 3} -{3\over 16} \Im(\epsilon^2)T \right), \qquad T\equiv t_f-t_i \gg 1.
\label{e7a}
\end{equation}
One can check that, for large $T$, solutions on the branches with positive $n$ are relatively suppressed by this exponent compared to the solution on the principal branch. However, for negative $n$, the solutions are relatively unsuppressed. These solutions have  an energy with a {\it positive} real part and presumably represent ``transients" due to the fact that the initial Gaussian wavefunction is not the true local metastable vacuum but includes an admixture of excited states. If, instead of using an initial Gaussian wavefunction, one uses the semiclassical local ground state wavefunction (calculated, for example, in Ref.~\cite{BenderWu}), one finds only one classical solution, as shall be reported elsewhere.  

In order to better understand the real-time solutions and compare them with the Euclidean bounce, we consider their behavior in the complex $t$-plane, shown in Fig. 2. As we have already mentioned, each solution is doubly periodic in $t$, with the fundamental domain consisting of a parallelogram with sides $a\equiv 2K\sqrt{1+m}$ and $b\equiv iK'\sqrt{1+m}$. Two corners of the parallelogram (shown as red points) are poles, and the other two corners (shown as green points) are zeros. The argument of the real-time solution  runs from some large negative $t_i$ towards $t_f$ which, without loss of generality, we can take to be zero. 

\begin{center}
\includegraphics[height=4in] {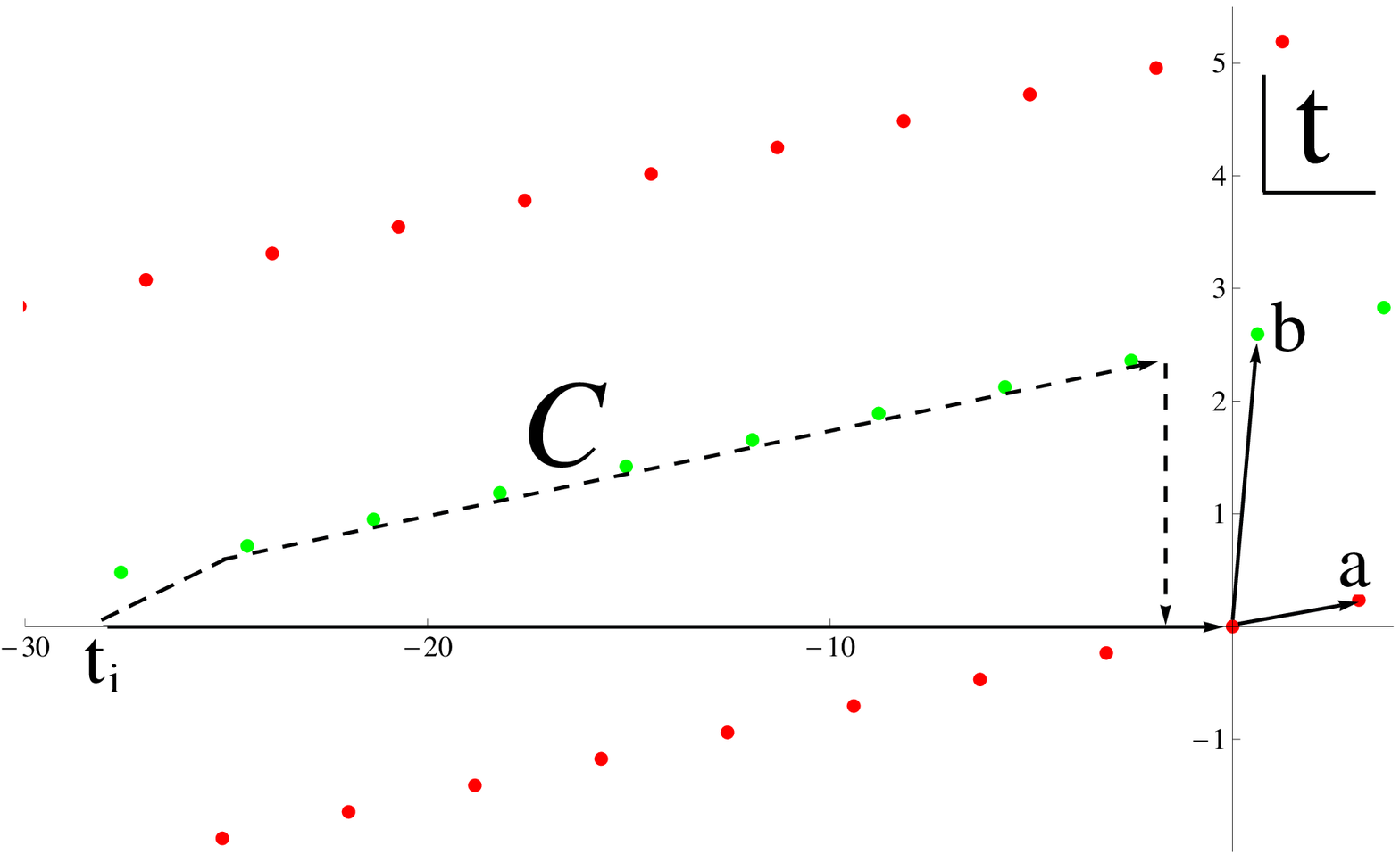}
\label{t2.eps}%
\end{center}
\begin{footnotesize}
Figure 2. Zeros and poles of the complex classical tunneling solution (\ref{e4}) are plotted in the complex $t$-plane. The plot shows the location of zeros and poles for  $m=i/10$ for  visual clarity, and for $t_0=0$: as the post-selected tunneling time $T$ tends to infinity, $\Im(a)$ becomes small and  $\Im (b)$ diverges so the lines of zeros and poles become more nearly horizontal, and the spacing between them grows to infinity. The solutions of interest, which start out near the metastable vacuum and end up running to infinite $x$, must, in this complex $t-$plane, start out near a zero  and end at a pole. 
\end{footnotesize}

Fig. 2 reveals an important property of the solution in the limit of long tunneling time $T$, or small $m$. Namely,  as we advance the solution in real time along the negative $t$-axis towards the origin, we pass close - a distance of order $m$ - to the penultimate pole before the one finally causing it to diverge as it runs to large $x_f$. This has the consequence that, for large $T$, the imaginary part of the solution becomes very large and negative - of order $-0.4/\epsilon \sim -T/\ln T$ at a time equal to one-quarter of the false-vacuum oscillation period before it emerges from the potential. We shall return to this later, when we discuss how to weakly measure the imaginary part of the solution.

Fig. 2 also allows us to relate our solution to the Euclidean instanton. Since the line of zeros above the real axis tilts downward and approaches the real $t$-axis as $t$ runs negative, in our solution $x$ becomes small at early times. As discussed earlier, at early times the behavior is exponential, $x\sim e^{{3\over 4} \epsilon t}$. In the semiclassical approximation, the wavefunction is proportional to the exponential of $iS/\hbar$ plus the `boundary term' coming from the initial wavefunction:
\begin{equation}
\Psi\sim e^{( \kappa^{3\over 2} m^{1\over 2}/ \hbar \lambda)\left( i\int_{t_i}^{t_f} dt {1\over 2} (\dot{x}_{Cl}^2-V(x_{Cl}))-{1\over 2} x_{Cl,i}^2\right)},
\label{e8}
\end{equation}
where $x_{Cl}$ is the classical solution. The classical action involves an integral along the real $t$-axis. Since the integrand is analytic away from its poles, the integration contour may be deformed to the contour marked $C$ in Fig. 2,  running along the line of zeros above the real $t$-axis, where the integrand is small. The deformed contour then has to return to $t_f$, on the real axis.  Since $b$ diverges logarithmically as $m$ tends to zero, the last part of the contour runs downwards from a large positive imaginary time. Taking the limit $m\rightarrow 0$, our solution (\ref{e4}), with $t_0=0$, tends to $x_{Cl}(t) \approx -1/\sin t$: setting 
$t=-\pi/2-i \tau$, we see that (\ref{e4}) becomes the Euclidean instanton, and the action, integrated over $-\infty < \tau<0$, reproduces the Euclidean result. Along the final part of the contour, from $-\pi/2 <t-t_0<t_f$,  the solution is very nearly real hence the corresponding action is imaginary and only contributes a phase to the wavefunction.

Thus, when the post-selected tunneling time $T\gg1$, we reproduce the Euclidean result for the exponent of the tunneling probability. However, our approach  is more versatile in several respects. First, since the initial wavefunction appears explicitly, it is easy to change it, for example with a shift in its centre, or width, or initial momentum. Each of these will change the initial condition in (\ref{e3}) and hence the corresponding classical solution. Likewise, it is straightforward to consider an initial wavefunction of a different shape. Finally, within this approach, it is simple to introduce time-dependence into the action, for example through a time-dependent forcing term, or, in the case of quantum fields, the expansion of the universe, in order to study their  effect on the tunneling rate.

\begin{center}
\includegraphics[height=5in] {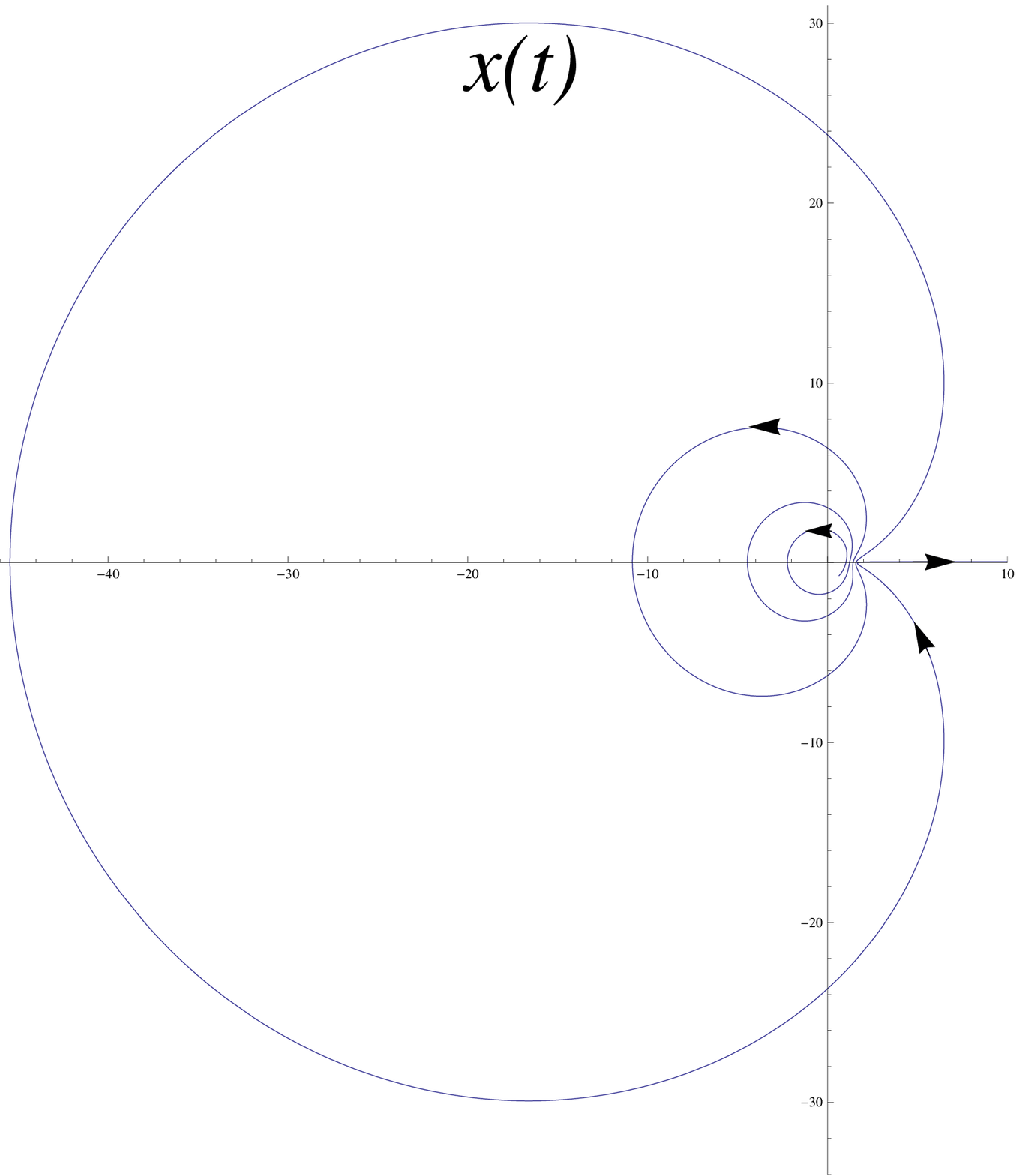}
\label{tplot2.eps}%
\end{center}
\begin{footnotesize}
Figure 3.  For the potential  $V(x)= {1\over 2} x^2-{1\over 3} x^3$, a complex classical solution $x(t)$  satisfying the boundary conditions given in (\ref{e3}) with $t_i=-30$, $t_f=0$, $x_f=\infty$   is plotted for real times $t_i <t <t_f$  in the complex $x$-plane.  The energy of this solution is $.046+.048 i$.
\end{footnotesize}

As a second instructive example, consider  a cubic potential, $V(x)={1\over 2} \kappa x^2 -{1\over 3} \lambda x^3$. In dimensionless coordinates for
$x$ and $t$, the action becomes $S =( \kappa^{5\over 2} m^{1\over 2}/ \lambda^2) \int dt ({1\over 2} \dot{x}^2-{1\over 2} x^2-{1\over 3} x^3)$. The general classical solution now has double poles rather than single poles in $t$: it may be written
\begin{equation}
x(t)=A+ {B\over {\rm sn}^2( C(t_0-t) |m)}, \quad E={1\over 12}-{(2m-1)(m-2)(m+1)\over 24(1+m(m-1))^{3\over 2}},
\label{e9}
\end{equation}
where $A={1\over 2}(1-{1\over \sqrt{1+m(m-1)}})$, $B={3\over 2 \sqrt{1+m(m-1)}}$, and $C={1\over 2 \sqrt[4]{1+m(m-1)}}$.

As before, we can express the solution using the Lambert series (\ref{e5}) and, at early times, expand it in terms of $m$ and $Z^{-1}$. We obtain
\begin{equation}
x(t)\approx -6 Z^{-2}(1+m)  -12 Z^{-4} -{3\over 128} m^2 Z^2+{63\over 64} m^2,
\label{e10}
\end{equation}
plus terms higher order in $m$ and $Z^{-1}$,  where $Z=e^{i U}$ and $U={\pi\over 2K} C(t_0-t) \approx {1\over 2} (1-{15\over 64} m^2)(t_0-t))+\dots$. As before, we impose the initial condition (\ref{e3}) at the initial time $t_i$, noting that  $x+i\dot{x}\approx-{3\over 64} m^2 Z^2+12 Z^{-4}$ at leading order, so that $m\approx 16 Z^{-3}$. Since $Z$ is of order $m^{-{1\over 3}}$, corrections to the expansion in (\ref{e10}) and in the initial condition are down by $O(m^{2\over 3})$ relative to the terms which are kept.  Finally, setting $m^2\equiv i \epsilon$, we find the transcendental equation for $\epsilon$
\begin{equation}
{45\over 64}  \epsilon T e^{{45\over 64}  \epsilon T}\approx -180 iT e^{-3 i T}.
\label{e11}
\end{equation}
Again, for large $T$, this equation is solved by the Lambert W function. The discussion parallels that preceding Eq. (\ref{e7a}) above. As mentioned, for large post-selected tunneling time $T$, the solution starts out small and grows exponentially, with a small exponent. Comparing the cubic and quartic cases, we find $|x|\sim e^{{\ln T\over n T} t}$ with $n=3$ and $4$ respectively.

Just as we did for the quartic potential, we can also determine the time at which the imaginary part of $x$ attains its greatest magnitude, just before the particle tunnels. In the cubic case, for small $m$, the argument of the elliptic function is approximately $t/2$, so the double pole to the left of the origin is located just below $t=-2 \pi$. The imaginary part in this case attains its greatest magnitude,  which is of order $1.5 \epsilon^{-2} \approx .75 (T/\ln T)^{2}$, one half-period of oscillation before the particle emerges from the potential and rolls downhill to large $x$. Since the pole is a double one, the imaginary part attains both very large positive and very large negative values in this case. 

These solutions provide a precise answer to the questions posed earlier: in the semiclassical limit, {\it a particle quantum tunnels by going around the potential barrier in the complex $x$-plane}.  It will be noticed from Fig. 1 and Fig. 3 that the trajectory actually approaches the `outside' of the barrier from the right and nearly comes to rest before bouncing off the potential and rolling down to large $x_f$. In the limit of large post-selected tunneling time $T$, the imaginary part of the particle's coordinate attains a magnitude of order $T/\ln T$ for a quartic potential and $(T/\ln T)^{2}$ for a cubic potential. As we have explained, the blow-up of the imaginary part is related to the singularities possessed by the classical solution in the complex $t$-plane. For more general potentials, such as those having a true minimum, one can expect the value of  $x$ at the potential minimum to play a role in determining the maximum value of the imaginary part. 

 It is very interesting to ask whether the complex nature of the classical tunneling solution, and in particular the large imaginary part it achieves just prior to tunneling, might be observable. As we shall now explain, within the framework of weak measurement, the answer is yes.

Just as Aharonov's concept of post-selection is useful in identifying the complex classical solution responsible for finite-time tunneling, his concept of weak measurement is useful in understanding what this solution means. The basic idea is that we would like to understand where the tunneling particle is, but without performing a measurement which would significantly disturb the particle's wavefunction. As Aharonov pointed out, we can do that by performing a weak measurement, which in our case means allowing the particle to interact weakly with a measuring device, and then performing the measurement  many times on a large ensemble of identically-prepared systems to obtain an accurate average. 

In order to measure the position $x$ of the particle, we introduce a simple von Neumann `pointer' Hamiltonian of the form
\begin{equation}
{\cal H}_P= {P^2\over 2 M} +g P\, x\, \delta(t-t_m),
\label{e12}
\end{equation}
where $P$ and $M$ are the pointer momentum and mass, $t_m$ is the measurement time and $g$ is a dimensionless coupling which is taken to be so small that the effect of the measurement on the system  being measured is negligible. Classically, the effect of this Hamiltonian is to shift the pointer position by the amount $g x(t_m)$ whereas the pointer momentum $P$ is conserved as a consequence of translation invariance and Noether's theorem. Thus, from the change in the pointer's position due to the measurement, one can infer the position of the particle at time $t_m$.  The pointer-particle interaction Hamiltonian also alters the momentum of the particle by $-g P$, but this influence becomes negligible for small $g$. Notice also that the pointer's kinetic term will be unimportant in what follows: one can take the limit $M\rightarrow \infty$ at the start and completely ignore it. In fact,  the discussion will be virtually unchanged if the term coupling the pointer to the tunneling coordinate involves not the pointer's momentum $P$ but its coordinate $X$ or even any linear combination of $P$ and $X$. In what follows we shall assume the coupling given in (\ref{e12}) but, in fact, the derivation  applies {\it mutatis mutandis} for any phase space coordinate $P$ and its conjugate coordinate $X$.

Quantum mechanically, since the pointer momentum $P$ is exactly conserved, the effect of the interaction is most easily seen in the $P$-representation. The total wavefunction may be represented as a  path integral: 
\begin{equation}
\Psi(x_f,P_f,t_f)={\cal N}  \int Dx dx_i \, DX DP dP_ie^{{i\over \hbar}\int_{t_i,x_i,P_i}^{t_f,x_f,P_f} \left( {1\over 2} \dot{x}^2 -V(x) -X \dot{P} -{\cal H}_P\right)}\Psi(x_i,t_i)\Psi_P(P_i,t_i),
\label{e13}
\end{equation}
where the initial pointer wavefunction $\Psi_P(P_i,t_i)\propto e^{-P_i^2 \Delta X^2 /\hbar^2}$. Here, $X$ is the position of the pointer and $\Delta X$ is its initial uncertainty. It is convenient to to use the pointer action in first order form in order to correctly implement the initial and final condition on the pointer momentum. To find the saddle point of (\ref{e13}), we first vary with respect to $X$ and obtain the classical equation of motion $\dot{P}=0$. So the pointer momentum is precisely its final value $P_f$ throughout, and the first term $-X\dot{P}$  in the pointer action gives no contribution. The $P^2/2M$ term, if present, contributes a phase which is irrelevant to the final probability. In the semiclassical approximation, and to first order in $g$, the pointer-particle interaction contributes a term  $ -ig P_f x_{Cl}(t_m)/\hbar$ in the exponent, where $x_{Cl}(t_m)$ is the original complex classical path.  The pointer-particle coupling alters the particle trajectory at order $g$ but this  only leads to corrections in the exponent which are of order $g^2$ and hence negligible at small $g$. 

By squaring the wavefunction, one finds that at leading order in $g$ the  effect on the average pointer position and momentum amounts to
\begin{equation}
\langle X \rangle \rightarrow \langle X \rangle+ g \Re(x_{Cl}(t_m)); \qquad \langle P \rangle \rightarrow \langle P \rangle+ {g\hbar \over 2 (\Delta X)^2}  \Im (x_{Cl}(t_m)),
\end{equation}
 where $x_{Cl}(t)$ is the original classical tunneling solution (see also \cite{ReschSteinberg}). Notice that, {\it even though the pointer momentum is strictly conserved, because the classical solution for the tunneling particle is complex, the action contributes a complex phase to the pointer wavefunction and this  produces a `quantum post-selection bias' in the final pointer momentum}. The bias is proportional to $g\hbar $ times the imaginary part of the particle's position $x_{Cl}$ at the measurement time. 
 
While the post-selection bias in the pointer momentum is proportional to $\hbar$, if the measurement time is carefully correlated with the post-selection time, the effect can be arbitrarily large in the $\hbar \rightarrow 0$ limit. This is because $ \Im(x_{Cl})$ can, in principle, have an even stronger dependence on $\hbar$.  As we have explained, for the quartic potential, if the particle coordinate is weakly measured a quarter period before the particle emerges from the well, its imaginary part attains a value of order $T/\ln T$ where $T$ is the post-selected tunneling time measured in the false vacuum oscillation period. The typical tunneling time is set by the tunneling exponent, $T_{typical} \sim e^{+S_E/\hbar}$, where $S_E$ is the classical Euclidean action, which is  independent of $\hbar$. It follows that the quantum post-selection bias in the pointer momentum is proportional $e^{S_E \over \hbar}$ times a power of $\hbar$ so that, in the formal limit $\hbar\rightarrow 0$, the effect actually diverges! A similar result holds in the cubic potential.
 
 It would be very interesting to perform a weak measurement on a tunneling particle and to observe the resulting post-selection bias in the pointer momentum. This would provide direct confirmation that, as far as its influence on the pointer is concerned, a post-selected particle follows a complex classical path. To see the large imaginary part requires a time-resolution better than the false vacuum oscillation period. For atoms or nuclei this is well beyond current experimental capabilities. However, other systems appear more promising. 
Tunneling can be observed in superconducting circuits which can be manipulated and probed on  timescales of order $10^{-10}$ seconds, comparable to their oscillation frequencies \cite{girvin, wilson}. Even more promising, in quantum dots, gate voltages can be changed in picosecond time-scales, much shorter than the characteristic oscillation periods or tunneling times.\cite{taylor}  Finally, for atoms tunneling in optical potentials~\cite{review},  the oscillation frequencies are even longer - of order microseconds.  The `quantum Zeno' effect has been observed in these systems but weak measurements of the type discussed here have not, as far as I am aware, been attempted.  With the current rapid progress in experimental capabilities,  we may well soon be able to directly confirm that, in quantum tunneling and after postselection, the system may be approximated by a classical  one following complex rather than real trajectories. 

The  calculational technique proposed here,  although technically intricate, is conceptually straightforward. Many possible developments and applications may be anticipated. Numerical approaches to mixed boundary value problems of the type needed here are in development,  and it is of great interest to extend them to cases where there are more degrees of freedom and/or explicit time-dependence.\cite{smithturok}  In this Letter I have focused only on the computation of the semiclassical exponent in the wavefunction, which is of order $1/\hbar$. The prefactor should be calculable in an expansion in $\hbar$. The zeroth order term is a functional determinant, which may be expressed in terms of the complex classical solution $x_{Cl}(t)$, as will be reported elsewhere. The method investigated here may also be extended to quantum field theory, in order to describe bubble nucleation in Minkowski spacetime for a theory with a metastable false vacuum state~\cite{Coleman:1977py} . One of the deficiencies of the Euclidean, imaginary-time formulation in that context is that there is no `two-instanton' solution and hence it is hard to describe rigorously  the nucleation of more than one bubble. The real-time formulation given here resolves this problem. Our method may also be extended to bubble nucleation in quantum field theories coupled to gravity \cite{Coleman:1980aw}, and is relevant to the resolution of certain paradoxes involving bubbles in the context of the `inflationary multiverse' scenario for the global structure of the universe. Finally, it is hoped that this method may allow a precise semi-classical, real-time description of the Hawking evaporation of black holes. These ideas and investigations will be reported elsewhere. 

I would like to thank Yakir Aharonov in particular for encouraging me to pursue this topic. Many thanks also to Asimina Arvanitaki, Jacob Barnett, Carl Bender, Michael Berry, Latham Boyle, Cliff Burgess, Curtis Callan, David Cory, Bianca Dittrich, Erik Schnetter, Davide Gaiotto, Steve Girvin, Lucien Hardy, Maxim Kontsevich, Sandu Popescu and Kendrick Smith as well as participants in the Aharonov 80th Birthday Conference \cite{ah80} for stimulating discussions.  I am specially grateful to Jacob Taylor for suggesting an experimental test of this phenomenon using quantum dots. Research at Perimeter Institute is supported by the Government of Canada through Industry Canada and
by the Province of Ontario through the Ministry of Research and Innovation.

\end{document}